\documentclass[conference]{IEEEtran}
\usepackage{verbatim}
\usepackage{amsfonts,amsmath,mathrsfs,amssymb,amsbsy}
\usepackage[final]{graphicx}
\usepackage{times,cite}
\usepackage{caption}
\usepackage{subcaption}

\long\def\symbolfootnote[#1]#2{\begingroup%
\def\thefootnote{\fnsymbol{footnote}}\footnote[#1]{#2}\endgroup}

\usepackage{verbatim}
\usepackage[]{float,latexsym}
\usepackage{amsfonts,amsmath,mathrsfs,amssymb,amsbsy}
\usepackage{url}

\newtheorem{theorem}{Theorem}[section]

\newcommand{\Prob}{\mathsf{P}}
\newcommand{\Expect}{\mathsf{E}}

\usepackage{color}
\definecolor{lightblue}{rgb}{.7, .8, 1}
\definecolor{lightgreen}{rgb}{.6, 1, .6}
\usepackage{color}
\definecolor{brown}{rgb}{1,0.38,0.03}

\definecolor{OliveGreen}{rgb}{.2,0.6,0.2}
\definecolor{BrickRed}{rgb}{.7,0.2,0.2}

\newcommand{\ignore}[1]{} 

\long\def\symbolfootnote[#1]#2{\begingroup%
\def\thefootnote{\fnsymbol{footnote}}\footnote[#1]{#2}\endgroup}

\newcommand{\bsp}{\begin{split}}

\newcommand{\esp}{\end{split}}

\begin{document}
%
\title{Sequential Event Detection Using Multimodal Data in Nonstationary Environments}

\author{\IEEEauthorblockN{
Taposh Banerjee\IEEEauthorrefmark{1},
Gene Whipps\IEEEauthorrefmark{2},
Prudhvi Gurram\IEEEauthorrefmark{3}\IEEEauthorrefmark{2},  
Vahid Tarokh\IEEEauthorrefmark{4}
}
\IEEEauthorblockA{\IEEEauthorrefmark{1}Harvard University, School of Engineering and Applied Sciences, Cambridge, MA 02138}
\IEEEauthorblockA{\IEEEauthorrefmark{2}U.S. Army Research Laboratory, Adelphi, MD 20783}
\IEEEauthorblockA{\IEEEauthorrefmark{3}Booz Allen Hamilton, McLean, VA 22102}
\IEEEauthorblockA{\IEEEauthorrefmark{4}Duke University, Department of ECE, Durham, NC 27708}
}


\maketitle

\begin{abstract}
The problem of sequential detection of anomalies in multimodal data is considered. 
The objective is to observe physical sensor data from CCTV cameras, and social media 
data from Twitter and Instagram to detect anomalous behaviors or events. Data from each modality is 
transformed to discrete time count data by using an artificial neural network to obtain counts of objects in CCTV images and by counting the number of tweets
or Instagram posts in a geographical area. The anomaly detection problem is then formulated as a problem of quickest detection of changes in 
count statistics. The quickest detection problem is then solved using the framework of partially observable Markov decision 
processes (POMDP), and structural results on the optimal policy are obtained. The resulting optimal policy is then applied 
to real multimodal data collected from New York City around a 5K race to detect the race. The count data both 
before and after the change is found to be nonstationary in nature. The proposed mathematical approach to this problem provides 
a framework for event detection in such nonstationary environments and across multiple data modalities. 
\end{abstract}


%
\IEEEpeerreviewmaketitle

\section{Introduction}
Event detection has many real-world applications such as surveillance \cite{panda2017, lee2014}, border security using unattended ground sensors (UGS) \cite{szechtman2008}, crime hot-spot detection for law enforcement \cite{neill2007}, cyber-infrastructure monitoring \cite{mitchell2013}, real-time traffic monitoring \cite{dandrea2015}, and environmental and natural disaster monitoring \cite{dereszynski2012, sakaki2010}. We address the problem of real-time event detection for gathering tactical intelligence, which is critical for military and law-enforcement missions. For instance, in tactical scenarios, like cordon and search, there is a need for gathering real-time intelligence that can help Soldiers at the squad level gain situational understanding of a scene and to quickly make mission-oriented decisions. To help gain such actionable information, Soldiers may deploy a variety of sensors such as cameras for imagery and video. The squad may also have access to auxiliary information such as SIGINT, SPOT reports, Blue Force Tracking data, and local social network feeds. Currently, relevant information is processed and analyzed in a far-off rear position, but there can be significant delays in receiving important decisions at forward positions. Some of the automated decisions that could help the squad carry out the mission successfully include the detections and locations of enemy entities such as personnel and vehicles. The goal is to provide real-time threat indicators from all available information sources at the point of need. 

We tackle two fundamental questions that arise in such scenarios. First, we have to process and fuse the information from traditional physics-based sensing systems, such as video sensors, and non-traditional sensing systems, such as social networks, to provide indications and warnings. Second, we have to push the processing to the operational environment where the information is needed most, and hence there is a need for real-time event detection. 
Motivated by these two questions, in this paper, we consider the problem of real-time event or anomaly detection using multimodal data. 


To solve the complex multimodal event detection problem, we need to develop useful mathematical models as well as efficient algorithms and validate them on real-world data collected during tactical scenarios. However, access to such data for research is severely restricted. To overcome this, we instead use publicly available data sources as surrogates for tactical data sources. In the case of imagery, we use New York City (NYC) traffic CCTV cameras as surrogates for low altitude UAVs with video sensors.  The video sensors onboard tactical UAVs typically are low-resolution and have a wide range of ground sample distances.  Simiar image qualities are present in the publicly available CCTV traffic camera imagery. Instagram has medium to high-resolution imagery, which can be viewed as surrogates for imagery collected by Soldier-worn cameras. In scenarios where social media posts are not available, the social media posts in this data collection could be viewed as surrogates for SIGINT data (e.g., counts of communications packets through local nodes). 

In this work, we are interested in the subtle information available in the dynamics of sequences of sub-events, e.g., changes in the counts of persons and vehicles in a spatial region and changes in the corresponding social network posts in the same region. 
As a result, we utilize the images from the CCTV cameras to extract counts of persons and vehicles in a spatial region. We also utilize the social media posts to generate count sequences of Twitter and Instagram posts in the constrained region.  
We develop a theoretical framework and a novel algorithm for sequential detection of changes in count statistics. The developed algorithm
is then applied to data collected from the NYC CCTV cameras and social media feeds to detect a 5K race. 
The proposed mathematical framework, and the developed algorithm can also be adapted to other event detection problems. For example, in cyber-infrastructure monitoring, the types and counts of intrusion attempts can indicate the onset of a coordinated attack \cite{harang2017}.

Towards developing a mathematical model for the problem, we first study the statistical behavior of the count data on the day of the event (a 5K race in NYC) and also on the non-event days (see Section~\ref{sec:DataCollAndMod}). We observe that the count sequences have nonstationary rates, i.e., the average counts of persons, vehicles, or social media posts, change over time, on each day.  Thus, the event detection problem of interest in this paper is a problem of detecting changes in the levels of nonstationarity of rates. 
We use the framework of POMDP to model the rate level change detection problem as detection of time 
to absorption in a hidden Markov model (HMM) \cite{krishnamurthy2016partially}, \cite{bertsekasshreve1978} (see Section~\ref{sec:ProbForm}).  
Our POMDP problem is more general than the one studied in \cite{krishnamurthy2011bayesian} as we detect both increases and decreases in rates. 
As a result, it is not apparent if our POMDP solution has the threshold structure that was found in the problem in \cite{krishnamurthy2011bayesian}. 
However, in this paper, we show that under certain assumptions on the transition structure of the HMM, the solution to our POMDP problem also has a simple threshold structure (see Section~\ref{sec:StructureOfOptPol}). 
We then apply the resulting belief sum algorithm to detect the event (see Section~\ref{sec:Numerical}).

\section{Data Collection and Modeling}\label{sec:DataCollAndMod}
We collected imagery from CCTV cameras and social networks around the Tunnel to Towers 5K run that occurred on September 24th, 2017, in NYC. Data was also collected on two weekends before the run, on September 10th and 17th, and a weekend after the run, on October 1st. CCTV imagery and social media posts were collected over a geographic region from the Red Hook village in Brooklyn on the south end to the Tribeca village on the north end of the collection area.  Data were collected between 8:30 am and 2 pm local on each of the 4 days.  On average, the frame rate from 7 CCTV cameras was roughly 0.5 frames per second.  While the average post rates from Twitter and Instagram for the geographic region and collection period were 1.4 and 0.7 posts per second, respectively.  
Note that for this initial modeling and analysis work, no other filtering of social posts was applied (e.g., hashtag clustering or content analysis).

\begin{figure}[ht]
	\center
	\includegraphics[width=9.5cm, height=6cm]{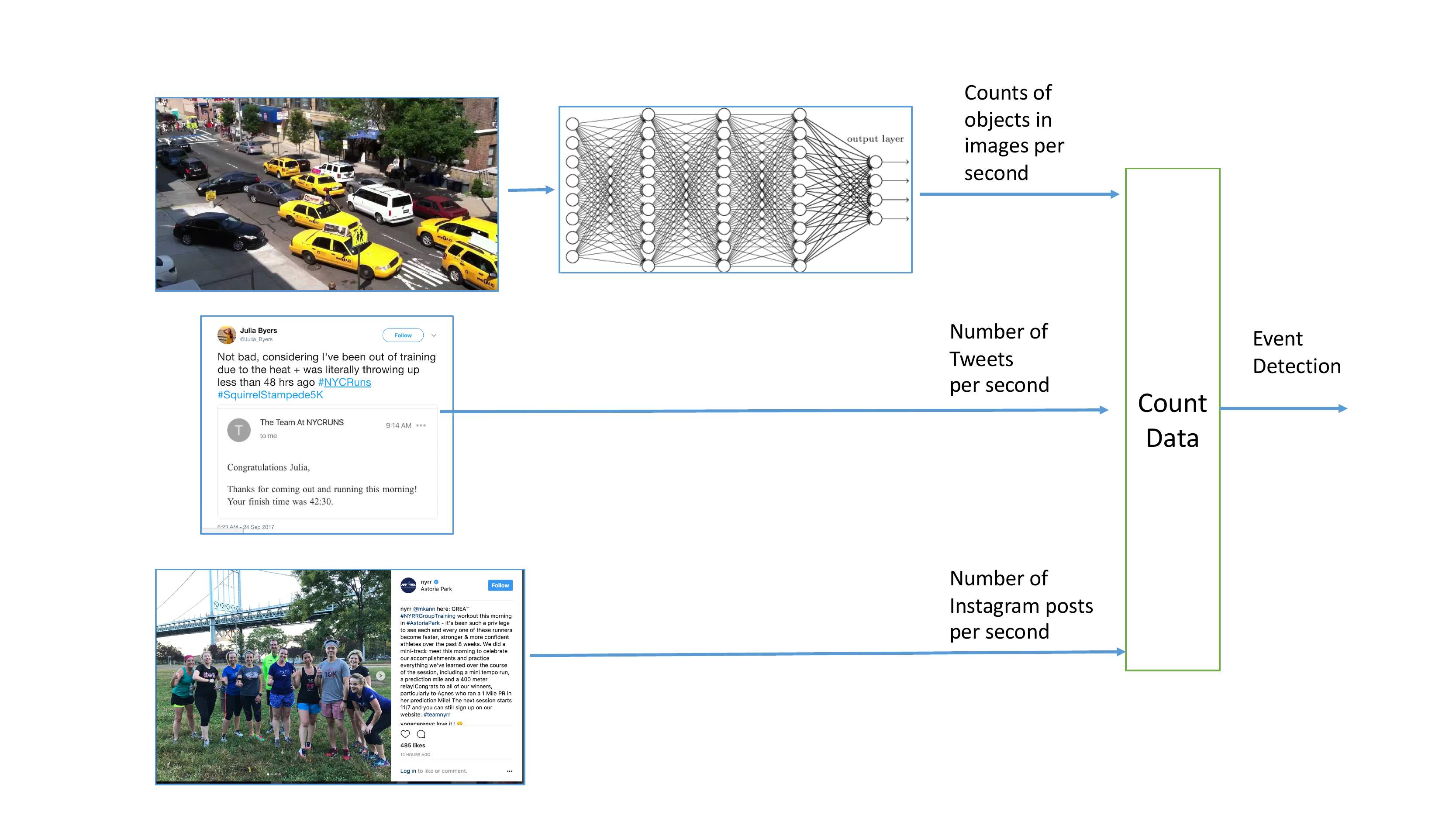}
	\caption{Event detection system}
		\label{fig:TransToCount}
\end{figure}

The objective is to detect the event in terms of location and time of the 5K run from the multimodal data. It is to be expected that the run would increase the number of persons on the streets overlapping with the route followed for the run. The run would also cause a sudden decrease in the number of cars on the same streets. It is also expected that the event would cause a surge in the number of tweets or Instagram posts pertaining to the event. Motivated by these observations, we approach this problem through the framework of quickest detection in count data. The multimodal data is used to obtain counts of objects (persons and vehicles) per frame and counts of tweets and Instagram posts per second. Fig.~\ref{fig:TransToCount} illustrates the block diagram of the event detection system.   

To obtain the counts of persons and vehicles, first, we use a Convolutional Neural Network (CNN) based object detector to detect persons and vehicles in each frame coming from the CCTV cameras. Specifically, we use faster R-CNN \cite{fasterrcnn2015}, which uses VGG16 architecture \cite{vgg162014} as the base CNN with region proposal networks, to perform real-time object detection. The faster R-CNN is trained on the PASCAL Visual Object Classes (VOC) dataset \cite{Everingham10}. The dataset has labeled training data for person class as well as vehicle classes that include bus, car, and motorbike. The counts of persons and vehicles are generated by simply counting the number of detected objects belonging to the corresponding class in each frame. 

\begin{figure}[ht]
	\center
	\includegraphics[width=8.5cm, height=5cm]{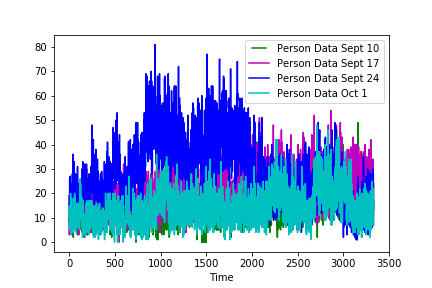}
	\caption{Total person counts from seven cameras for each day. Horizontal axis is time in multiple of six seconds.}
	\label{fig:CountAllPersons}	
\end{figure}

\begin{figure}[ht]
	\center
	\includegraphics[width=8.5cm, height=5cm]{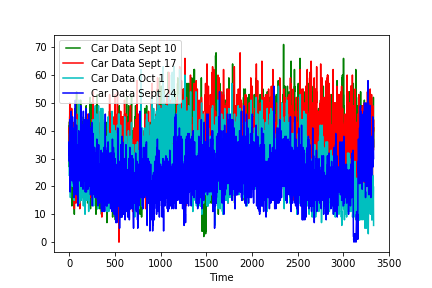}
	\caption{Total person counts from seven cameras for each day.}
\label{fig:CountAllCars}	
\end{figure}

\begin{figure}[ht]
	\center
	\includegraphics[width=8.5cm, height=5cm]{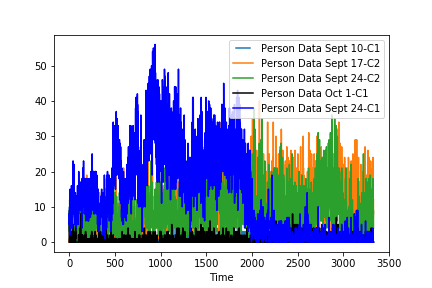}
	\caption{Person counts from two specific camera C1 and C2 on different days. Camera C1 is on the path of the race, while camera C2 is off the path.}
	\label{fig:CountPerCameraPersons}	
\end{figure}

\begin{figure}[ht]
	\center
	\includegraphics[width=8.5cm, height=5cm]{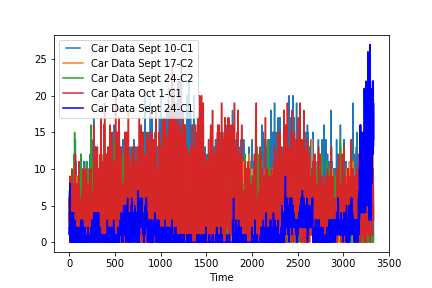}
	\caption{Car counts from two specific camera C1 and C2 on different days.}
	\label{fig:CountPerCameraCars}	
\end{figure}

\begin{figure}[ht]
	\center
	\includegraphics[width=8.5cm, height=5cm]{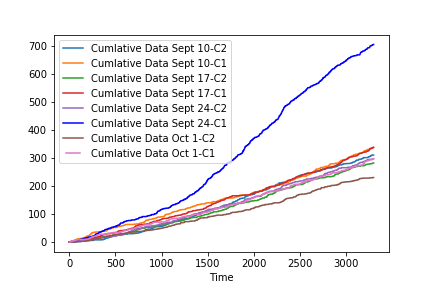}
	\caption{Cumulative counts of Instagram posts in geographical areas near cameras C1 and C2. }
\label{fig:CountInstagram}	
\end{figure}

In Fig.~\ref{fig:CountAllPersons}, we have plotted the total person counts, summed across the seven CCTV cameras of interest, for each of the four separate dates. Similar data for the total number of cars are shown in Fig.~\ref{fig:CountAllCars}. 
In this figure and the figures in the rest of this paper, the horizontal axis is time  in multiples of six seconds. 
In Fig.~\ref{fig:CountPerCameraPersons}, we have plotted the person counts for two cameras: camera C1 is on the path of the race while camera C2 is off the path. In Fig.~\ref{fig:CountPerCameraCars}, we have similarly plotted individual car counts for camera C1 and C2. We see a clear increase in the rate of the number of persons and a slight decrease in the number of cars on the day of the event between the time slots $500$ and $2000$. Finally, in Fig.~\ref{fig:CountInstagram}, we
have plotted cumulative counts of Instagram posts in geographical vicinity of camera C1 and C2 for the four days. 
We see an increase in the cumulative Instagram counts just around time slot $1500$, just before the person and car counts return to their normal rates. We hypothesize that the latter is due to the fact that the participants started posting on social media after completing the race. 

From these figures, one can make an observation that the rates and counts are nonstationary in nature. Thus, 
the problem of event detection here can be posed as a detection problem in nonstationary environments. Since the event detection has to be performed in real time, this would translate to sequential detection of changes in rate from one nonstationary level to another. In the next section, we 
formulate this problem in a POMDP framework and solve it to obtain structural results on the optimal solution. The resulting optimal 
algorithm will then be applied to the collected data to detect the event. 

\section{Problem Formulation}\label{sec:ProbForm}
We note that the count data generated from any modality is a sequence of discrete positive integers. For simplicity, we model the count data as a 
sequence of Poisson random variables. The results below are also valid for any single parameter probability distribution, discrete or continuous, with sums replacing integrals, where appropriate. 
Also, for simplicity, we develop the theory for event detection in a single stream of count data here. The resulting algorithm is then trained and applied to every count sequence generated from every modality. However, the mathematical model can easily be extended to a vector stream of observations to detect an event jointly across modalities.  

As observed in Fig.~\ref{fig:CountAllPersons}--Fig.~\ref{fig:CountInstagram}, count sequences are nonstationary in nature, on both the 
event day and the no-event days. In order to capture this nonstationarity, we model the count data as an HMM. In this HMM, there are a finite number of hidden states, and for each hidden state, the rate or mean of the observed count or Poisson random variable is different. Thus, if $Y$ represents the observed count variable, $X$ represent the hidden state variable, and if $\lambda_1 < \cdots < \lambda_N$
are $N$ possible rates for pre-event data, then 
$$
Y | X=k \; \sim \; \text{Pois}(\lambda_k).
$$

When the event starts, which we call a change point, the rate of counts either decreases (as in the case of cars) or increases (as in the case of persons). In practice, the post-change rates may not be known or may be hard to learn due to a lack of enough training data of a rare event. Motivated by this, we model the pre-change and post-change rates by boundary rates $\lambda_0$ (to capture a decrease in rates) and $\lambda_{N+1}$ (to capture an increase in rates) such that 
\begin{equation*}
\lambda_0 < \lambda_1 < \cdots < \lambda_N < \lambda_{N+1}.
\end{equation*}
In other words, $\lambda_0$ and $\lambda_{N+1}$ represent the minimum amount of change the designer of the system is interested in detecting. 
Note that while the number of cars decreases during the event in this data, one may also observe an increase in the numbers due to congestion of traffic. Thus, both increase and decrease of rates are of interest to us, and our model allows for both these possibilities. 

We have $N+2$ states, $N$ normal states with corresponding Poisson rates $\{\lambda_1 < \cdots < \lambda_N\}$, and $2$ abnormal states corresponding to Poisson rates $\{\lambda_0, \lambda_{N+1}\}$. Our aim is to observe the Poisson count data $Y$, and optimally detect the change of the hidden rate from normal to abnormal rates. Specifically, we want to detect this change as quickly as possible subject to a constraint on the false alarm rate. This leads us to the realm of 
quickest change detection \cite{veer-bane-elsevierbook-2013}, \cite{poor-hadj-qcd-book-2009}, \cite{tart-niki-bass-2014}.
Here, we solve the rate change detection problem by formulating it as a POMDP \cite{krishnamurthy2016partially}, \cite{bertsekasshreve1978}. 

\subsection{POMDP Formulation}
\begin{itemize}
\item \textit{States}\\
Let $\{X_k\}$ be the sequence of states with values $\{x_k\}$. The state process is a finite state Markov chain taking values $X_k \in \{\mathcal{A}, 0, 1, 2, \cdots, N, N+1\}, \; \forall k$.
The state $\mathcal{A}$ is a special absorbing state introduced for mathematical convenience in a stopping time POMDP \cite{krishnamurthy2016partially}. Its role will be clear when we define the cost structure below. 
The transition probability matrix of the Markov chain is a function of the control, and will also be defined below. 
\medskip
\item \textit{Control}\\
The control sequence $U_k$ taking values $\{u_k\}$ is binary valued: $U_k \in \{1 \;(\text{stop}), 2 \;(\text{continue}) \}$.
The control $U_k=2$ is used to continue the observation process and $U_k=1$ is used to stop it. 
At the time of stopping, an alarm is raised indicating that a change in the rate of the data has occurred. 
\medskip
\item \textit{Observations}\\
The observations are Poisson distributed with rate $\lambda_m$, if the state $X_k=m \in \{0, \cdots, N+1\}$ and if the control is to continue:
\begin{equation*}
\begin{split}
(Y_k | X_k = m, &U_{k-1}=2) \sim \text{Pois}(\lambda_m), \\
&m \in \{0, 1, 2, \cdots, N+1\}, \; \forall k.
\end{split}
\end{equation*}
The distribution of observations, if the state is $\mathcal{A}$ or if the control is to stop, is irrelevant. 
We use $B_{xy}(u)$ to denote the law of $Y=y$ when the state is $X=x$ and the control is $U=u$.
We also assume 
that the variable $Y_k$ is independent of the past states and controls, given the current state and last control. That is
\begin{equation*}
\begin{split}
(Y_k | & X_k, U_{k-1}=2)  \sim \\
&(Y_k | X_1, \cdots, X_k, U_1, \cdots, U_{k-2}, U_{k-1}=2)
\end{split}
\end{equation*}
\medskip
\item \textit{Transition Structure}\\
The transition structure depends on the control $U_k$. Let $P_{k+1|k}(u_k)=P(u_k)$ be the transition matrix for the 
Markov chain from time $k$ to $k+1$, given the control is $U_k=u_k$. Then,we have
\begin{equation*}
P_{k+1|k}(u_k) =P(u_k)= \begin{cases}
               P_1, \quad \text{if } u_k = 1\\
               P_2, \quad \text{if } u_k = 2.
            \end{cases}
\end{equation*}
Here, 
\begin{equation*}
P_1 = \begin{bmatrix}
    1 &0 & 0 & \dots  & 0 \\
    1 & 0 & 0 & \dots  & 0 \\
    \vdots & \vdots & \vdots & \ddots & \vdots \\
   1 & 0 & 0 & \dots  &0
\end{bmatrix}
\end{equation*}
and
\begin{equation}\label{eq:P2}
P_2 = \begin{bmatrix}
    1 &0 & 0 & \dots  &0& 0 \\
    0 & a_1 & 0 & \dots  &0& 1-a_1 \\
    0 & p_{10}&p_{11}& \dots  &p_{1N}&p_{1(N+1)}\\
    \vdots & \vdots & \vdots & \ddots & \vdots & \vdots\\
        0 & p_{N0}&p_{N1}& \dots  &p_{NN}&p_{N(N+1)}\\
   0 & 1-a_{N+1} & 0 & \dots  &0&a_{N+1}
\end{bmatrix}.
\end{equation}
To understand these two transition structures, we first define the initial distribution $\pi_0$ for the Markov chain $\{X_k\}$ as
\begin{equation*}
\pi_0 = (\pi_0(\mathcal{A}), \pi_0(0), \pi_0(1), \cdots, \pi_0(N), \pi_0(N+1))^T,
\end{equation*}
which satisfies $\pi_0(\mathcal{A})=\pi_0(0)=\pi_0(N+1)=0$.
Thus, the Markov chain $\{X_k\}$ starts in one of the states $\{1, \cdots, N\}$. 
As long as the control $U_k=2$, which means to continue, the states evolve according
to the transition probability matrix $P_2$. The transition probabilities 
\begin{equation}\label{eq:Pbar}
\bar{P} = \begin{bmatrix}
    p_{10}&p_{11}& \dots  &p_{1N}&p_{1(N+1)}\\
    \vdots & \vdots & \ddots & \vdots &\vdots\\
        p_{N0}&p_{N1}& \dots  &p_{NN}&p_{N(N+1)}
\end{bmatrix}
\end{equation}
that are part of the matrix $P_2$ in \eqref{eq:P2} control the transition of the Markov chain within the states 
$\{1, \cdots, N\}$, and its jump to the absorbing states $0$ and $N+1$. We assume that absorption to 
the states $0$ and $N+1$ is inevitable. Once in these two states, the Markov chain jumps between these two 
states with probabilities controlled by $a_0$ and $a_{N+1}$. We are especially interested in the case when $a_1 = a_{N+1} = 1$. 
This is because the states $\{1, \cdots, N\}$ correspond to the normal states for the counts before the change. After the change, we expect that either the rate will increase, corresponding to absorption of the Markov chain to the state $N+1$, or it will decrease, corresponding to absorption to the state $0$. Once the rate increases or decreases, it is unnatural 
to expect that rate will transition between too low and too high rates. However, the case $a_1 = a_{N+1} = 0.5$,
is of mathematical interest, and its role and importance will be briefly discussed below. 

\medskip
\item \textit{Cost}\\
Our objective is to detect a change in the rate of counts from normal rates $(\lambda_1, \cdots, \lambda_N)$
to abnormal rates $\lambda_0$ and $\lambda_{N+1}$. This is equivalent to detecting the time to absorption of the Markov chain $\{X_k\}$ from 
the states $(1, \cdots, N)$ to the states $(0, N+1)$. We now define a cost structure for the POMDP to capture
the sequential event detection framework. Let $C(x,u)$ be the cost associated with state $X=x$, and control $U=u$ and defined as
\begin{equation*}
\begin{split}
C(x,1) &= C_f^T e_x = (0, 0, c_f, c_f, \cdots, c_f, 0) \; e_x,\\
C(x,2) &= C_d^T e_x = (0, c_d, 0, 0, \cdots, 0, c_d) \; e_x.
\end{split}
\end{equation*}
Here, $e_x$ is a unit column vector with value $1$ at the $x$th position. The constant $c_f$ captures the cost of false alarm, and is incurred when the control is to stop and the state is in $(1, \cdots, N)$. Similarly, $c_d$ captures the cost of delay, and is incurred when the control is to continue even if the Markov chain is absorbed in either of the states  $(0, N+1)$. Note that the cost of being in state $\mathcal{A}$ is zero independent of the choice of control.  

\medskip
\item \textit{Policy}\\
Let $I_k = (y_1, \cdots, y_k, u_1, \cdots, u_{k-1})$ be the information at time $k$. Also define a policy $\Phi = (\phi_1, \phi_2, \cdots)$ to be a sequence of mappings such that $u_k = \phi_k(I_k), \; \forall k$.
\end{itemize}

\medskip
We want to find a control policy so as to optimize the long term cost, which is
\begin{equation*}
V(\pi_0) = \min_\Phi \Expect\left[\sum_{k=1}^\infty C(x_k, u_k)\right], 
\end{equation*}
where $u_k = \phi_k(I_k)$. Let $\tau = \inf\{k: x_{k+1} = \mathcal{A}\}$.
Then, 
\begin{equation}\label{eq:Vpi}
V(\pi_0) = \min_\Phi \Expect\left[\sum_{k=1}^\tau C(x_k, u_k)\right] .
\end{equation}
Thus, the cost is finite if $\Expect[\tau] < \infty$. The role of the extra state $\mathcal{A}$ is now clear. After the stopping 
control $u_k=1$ is applied, the Markov chain's transition is governed by the transition matrix $P_1$. As a result, 
the Markov chain gets absorbed into the state $\mathcal{A}$ immediately. From here, due to the cost structure, 
the cost to go is zero, no matter what control is chosen. In conclusion, we search over policies $\Phi$ for which $\Expect[\tau] < \infty$. 
This is hardly a concern since any open-loop policy, where we always stop at a fixed time, satisfies this condition. We are looking for policies better than that, i.e., for closed-loop control that allows us to stop dynamically after observing the system. 

\section{Structure of the Optimal Policy}\label{sec:StructureOfOptPol}
Let $\pi_k$ be the belief at time $k$ defined as $\pi_k = \Prob(X_k = x_k | I_k)$.
Note that the belief is a vector of length $N+3$.
By standard Bayes arguments, this belief can be computed recursively as 
\begin{equation}\label{eq:belief}
\pi_{k+1} = T(\pi_k, y_{k+1}, u_k) =\frac{B_{y_{k+1}}(u_k) \; P(u_k)^T \pi_k}{\mathbf{1}^T B_{y_{k+1}}(u_k) \; P(u_k)^T \pi_k}.
\end{equation}
Here, $B_y(u)$ is a diagonal matrix of emission probabilities
$$
B_y = \text{diag}(B_{\mathcal{A}y}(u), B_{0y}(u), \cdots, B_{(N+1)y}(u))
$$
and $\mathbf{1} = (1, \cdots, 1)^T$
is a vector of all $1$s of length $N+3$.

\medskip
It is a standard result in the POMDP literature that the cost to go $V(\pi)$ in \eqref{eq:Vpi} satisfies
the Bellman's equation
\begin{equation*}
V(\pi) = \min\left\{ C_f^T \pi, \; C_d^T \pi \; + \; \sum_y V(T(\pi, y, 2)) \; \sigma(\pi, y, 2)\right\},
\end{equation*}
where $\sigma(\pi, y, u) = \mathbf{1}^T B_{y}(u) P(u)^T \pi$.
It can be shown that the optimal policy is stationary and is a function of only the belief state. Furthermore, the value function can be computed using 
value iteration \cite{krishnamurthy2016partially}, \cite{bertsekasshreve1978}. That is, the optimal policy is of the form $\Phi^* = (\mu^*(\pi), \mu^*(\pi), \cdots)$.
In addition, the following result can also be shown. 
\begin{theorem}[\cite{lovejoy1987convexity}, \cite{krishnamurthy2016partially}]
Let 
\[
R_1 = \{\pi: \mu^*(\pi) = 1\}
\]
be the region of the belief space on which the control $U=1$ is chosen, or the stopping decision is made. Then $R_1$ is convex.
\end{theorem}

\medskip
A standard approach to solving POMDP problems, which are typically hard to solve due to the high-dimensionality of the belief space, is to 
establish additional structural results on the policy $\mu^*$. Specifically, it is of interest to show that the optimal stopping time in a POMDP has a threshold structure, or the policy $\mu^*(\pi)$ is, in some sense, monotone in $\pi$. The threshold structure motivates the use of policies that are linear in the belief state. 
See \cite{krishnamurthy2016partially} for a detailed discussion. 

Unfortunately, all the conditions needed to establish the threshold structure are not satisfied in our problem. For example, the 
transition structure and the emission probabilities satisfy the so-called total positivity conditions. But, the cost structure does not have 
the required monotonicity and submodularity structure \cite{krishnamurthy2016partially}. Even a transformation of the problem, 
as suggested in \cite{krishnamurthy2016partially} does not help. The main issue is that in comparison with the results in \cite{krishnamurthy2011bayesian}, in this paper, we have two absorbing states, one for the low rate and another for the high rate. However, we now establish that under some additional assumptions, the optimal policy can be shown to be only a function of the probabilities $\pi_k(0)$ and $\pi_k(N+1)$.

\medskip
\begin{theorem}\label{thm:bothbeliefs}
Let the rows of the matrix $\bar{P}$ in \eqref{eq:Pbar}, and hence the corresponding elements of $P_2$ in \eqref{eq:P2}, be identical. 
Also, let $a_1=a_{N+1}=1$ in $P_2$. 
Then, the optimal policy depends only on the two components  $\pi(0)$ and $\pi(N+1)$ of the belief state $\pi$.
\end{theorem}
\begin{IEEEproof}
Note that even without the addition assumptions that are made in the theorem statement, the value function satisfies the condition
\begin{equation*}
\begin{split}
V(\pi) &= \min\left\{ C_f^T \pi, \; C_d^T \pi \; + \; \sum_y V(T(\pi, y, 2)) \; \sigma(\pi, y, 2)\right\}\\
        &= \min\left\{ c_f (1-\pi(0)-\pi(N+1)), \right. \\
        &\left. c_d (\pi(0)+\pi(N+1)) + \; \sum_y V(T(\pi, y, 2)) \; \sigma(\pi, y, 2)\right\}.
\end{split}
\end{equation*}
This is because of the special structure of the cost function assumed in the paper. 
Now, we can show that under the assumptions of the theorem, the belief recursion 
\eqref{eq:belief} can be computed just based on the values of  $\pi(0)$ and $\pi(N+1)$. Hence, the fixed point equation 
of the value function is only a function of these two values. The rest follows by using the standard value iteration arguments. 
\end{IEEEproof}

\medskip
The condition that the rows of $\bar{P}$ be same is easily satisfied in the following special case:
\begin{equation}\label{eq:barPsamerows}
\bar{P} = \begin{bmatrix}
    \frac{1}{N+2}&\frac{1}{N+2}& \dots  &\frac{1}{N+2}&\frac{1}{N+2}\\
    \vdots & \vdots & \ddots & \vdots &\vdots\\
        \frac{1}{N+2}&\frac{1}{N+2}& \dots  &\frac{1}{N+2}&\frac{1}{N+2}
\end{bmatrix}.
\end{equation}
Thus, with this choice of $\bar{P}$, the Markov chain moves around the states $1$ to $N$ randomly
and can get absorbed to $0$ and $N+1$, all with equal probability. Numerical evaluation of the optimal policy for this case, 
under some parameters of choice, shows that the optimal policy is a function of $\pi(0)$ and $\pi(N+1)$ only through $\pi(0)+\pi(N+1)$. 
In fact, according to this numerical study, the optimal stopping rule is 
\begin{equation}\label{eq:OptSumStatAlgo}
\tau^* = \inf\{k: \pi_k(0)+\pi_k(N+1) > A\}.
\end{equation}

We note that although the marginal costs are a function of $\pi(0)$ and $\pi(N+1)$ only through $\pi(0)+\pi(N+1)$, 
the belief recursion cannot be computed just using this sum. To compute \eqref{eq:belief} we need both $\pi(0)$ and $\pi(N+1)$ individually. 
This can be verified by explicitly writing the belief recursion under the stated assumptions. Thus, it is not clear to us
at this moment if this threshold structure, the optimal policy being only a function of the sum $\pi(0)+\pi(N+1)$, holds for more general cases.

However, if we make the assumption that $a_1=a_{N+1}=0.5$, then in this case, we can show that the optimal policy is only a function of the sum 
$\pi(0)+\pi(N+1)$. 
\medskip
\begin{theorem}\label{thm:sumbeliefs}
Let the rows of the matrix $\bar{P}$ in \eqref{eq:Pbar}, and hence the corresponding elements of $P_2$ in \eqref{eq:P2}, be identical. 
Also, let $a_1=a_{N+1}=0.5$ in $P_2$. 
Then, the optimal policy depends only on the two components  $\pi(0)$ and $\pi(N+1)$ of the belief state $\pi$ through their sum $\pi(0)+\pi(N+1)$.  
\end{theorem}
\begin{IEEEproof}
Under the stated assumptions, the belief recursion can be shown to be a function of only the sum $\pi(0)+\pi(N+1)$. The rest of the proof
 is identical to that of Theorem~\ref{thm:bothbeliefs}. 
\end{IEEEproof}

\medskip
Note that this means that under the assumptions made in Theorem~\ref{thm:sumbeliefs}, and in Theorem~\ref{thm:bothbeliefs}, the quickest change detection problem studied here reduces to the classical quickest 
change detection problem \cite{shir-opt-stop-book-1978} in some sense. 
In the classical problem, there are two hidden states, one before the change and 
another after the change. The hidden Markov chain starts with one state and gets absorbed into the other. The objective in the classical change point problem is to detect this time to absorption. The optimal algorithm for the classical problem is to stop the first time, the belief that
the Markov chain is in the post-change state, is above a threshold. 

In the change point problem considered in this paper, we have two classes of states, 
one class consisting of pre-change or normal states $\{1, \cdots, N\}$, and another class consisting of post-change or abnormal states $\{0, N+1\}$. 
And the objective here is to detect the time at which the Markov chain $\{X_k\}$ moves from the pre-change class to the post-change class. 
The above theorems suggest that under the stated assumptions, the optimal stopping rule has a similar structure.
That is, it is optimal to stop the first time the probability that the Markov chain is in the post-change class of states is above a threshold, such as that in stopping rule \eqref{eq:OptSumStatAlgo}. 

\section{The Belief SUM Algorithm and its Applications}\label{sec:Numerical}
In the previous section, we established conditions on the transition structure of the HMM under which 
the algorithm, 
\begin{equation}\label{eq:SumBelief_1}
\tau^* = \inf\{k: \pi_k(0)+\pi_k(N+1) > A\},
\end{equation}
is optimal. However, note that under more general cost structures, it is not obvious if this is still the optimal algorithm. 
As a result, and motivated by the optimality of \eqref{eq:SumBelief_1} under some assumptions, we use a more general class of algorithms that are linear in the beliefs $\pi_k(0)$ and $\pi_k(N+1)$. That is, we use the sum belief algorithm using a convex combination
of the beliefs
\begin{equation}\label{eq:SumBelief_2}
\tau^* = \inf\{k: \alpha \;\pi_k(0)+(1-\alpha)\;\pi_k(N+1) > A\},
\end{equation}
where $\alpha \in [0,1]$, and optimize over the choice of $\alpha$. 
Due to a paucity of space, a detailed delay and false alarm analysis of this algorithm will be reported elsewhere. 
In this section, however, we apply the algorithm to real data to show its effectiveness. In the following, 
we often use $\alpha=0.5$. In those cases, we actually report values of the sum statistic in \eqref{eq:SumBelief_1}.

\subsection{Global Event Detection}
We now apply the algorithm to data collected around the 5K run. The details on the data are provided in Section~\ref{sec:DataCollAndMod}. 
In practice, the algorithm should be applied to data collected from each individual source: to outputs of each camera and also to outputs of social media data in each geographical region. A high value of the sum statistic would indicate an abnormal behavior 
in a stream. This can be used to both detect and isolate the event. This is done in the next subsection. However, we may also wish to apply the algorithm to the global sum of data collected to detect global trends, in case they generate a collaborative effect. 

We applied the algorithm to total count data from all cameras for global detection of the event. 
We first used the data from the first recorded day, Sept. 10th, to learn the Poisson rates. 
We then applied the trained algorithm to data from other days. The parameters used were $\lambda_0=0.001$, $\lambda_1=5$, $\lambda_2=10$, $\lambda_3=15$, $\lambda_4=20$, $\lambda_5=25$, and $\lambda_6=65$, i.e., $N=5$. 
The transition matrix used was assumed fixed as in \eqref{eq:barPsamerows}
and \eqref{eq:P2} with $a_1 = a_{N+1}=1$. 
Note that
the rate parameters we learn from the data are $\lambda_1$ to $\lambda_N$. The values $\lambda_0$ and $\lambda_{N+1}$ are chosen
to be the boundary of the learned rates based on multiple standard deviations from normal rates. 

\begin{figure}[ht]
\center
	\includegraphics[width=7cm, height=5cm]{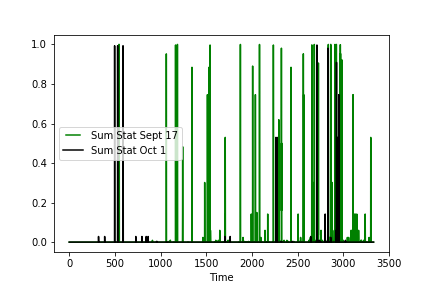}
		\caption{Evolution of the sum statistic \eqref{eq:SumBelief_1} for data from the non-event days Sept. 17 and Oct 1.}
	\label{fig:SumStatTotalPersonNonEvent}
	\end{figure}
\begin{figure}[ht]
	\center
	\includegraphics[width=7cm, height=5cm]{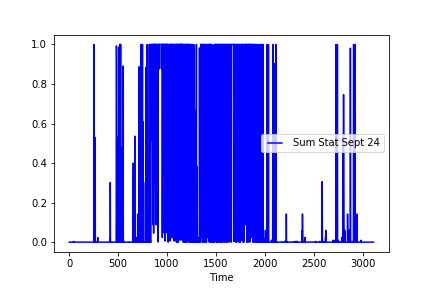}
		\caption{Evolution of the sum statistic \eqref{eq:SumBelief_1} for data from the event day Sept. 24th. We see high-values of the statistic between the times $500$ and $2000$, the times of the event as seen in Fig.~\ref{fig:CountAllPersons}.}
	\label{fig:SumStatTotalPersonEvent}
\end{figure}

In Fig.~\ref{fig:SumStatTotalPersonNonEvent} and Fig.~\ref{fig:SumStatTotalPersonEvent}, we report the results on application to total person count data. 
The data for the event day and a non-event day is shown in Fig.~\ref{fig:CountAllPersons}.  
Note that the statistic is sporadically large on the non-event days, but 
consistently fires around the event on the day of the event. 

\begin{figure}[ht]
	\center
	\includegraphics[width=7cm, height=5cm]{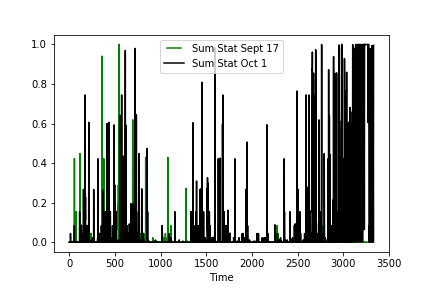}
		\caption{The evolution of the sum statistic \eqref{eq:SumBelief_1} for data from the non-event days. }
	\label{fig:SumStatCarsNonevent}
\end{figure}

\begin{figure}[ht]
	\center
	\includegraphics[width=7cm, height=5cm]{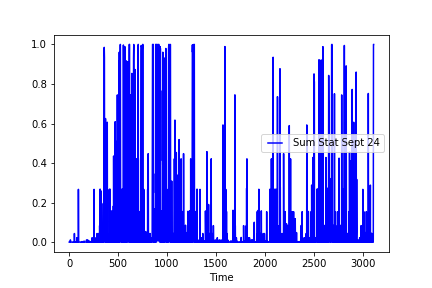}
		\caption{The evolution of the sum statistic \eqref{eq:SumBelief_1} for total car data from the event day. We see sustained high values for the statistic around the time of the event. }
	\label{fig:SumStatCarsEventDay}
\end{figure}

Similar results are reported in Fig.~\ref{fig:SumStatCarsNonevent} and Fig.~\ref{fig:SumStatCarsEventDay} for the total car count data 
from Fig.~\ref{fig:CountAllCars}. 
The learned values of rates from Sept. 10th data are $\lambda_0=0.00001$, $\lambda_1=0.001$, and $\lambda_2=55$, with the rest of the parameters kept the same. As seen in Fig.\ref{fig:SumStatCarsNonevent}, the statistic fires sometimes even on the non-event days. This is because we are applying the algorithm to the sum of count data from all cameras, and this may reduce the quality of the count sequence. 

\subsection{Event Localization}
\begin{figure}[ht]
	\center
	\includegraphics[width=7cm, height=5cm]{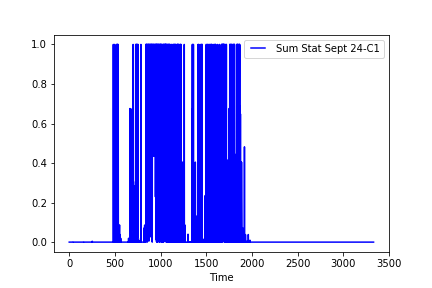}
		\caption{Evolution of the sum statistic \eqref{eq:SumBelief_2} with $\alpha=0$ applied to data from camera C1 on the day of the event. }
	\label{fig:SumStatPersonEventCamera}
\end{figure}

\begin{figure}[ht]
	\center
	\includegraphics[width=7cm, height=5cm]{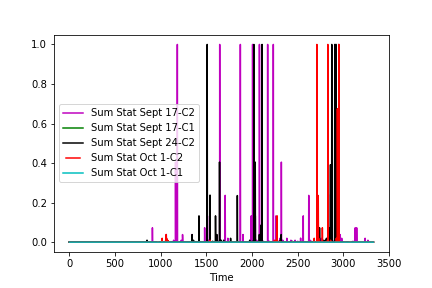}
		\caption{Evolution of the sum statistic \eqref{eq:SumBelief_2} with $\alpha=0$ applied to data from camera C1 on non-event days and camera C2, which is off the run path, on event and non-event days.}
	\label{fig:SumStatPersonNonEventCameras}
\end{figure}

In this section, we report results on the application of the algorithm \eqref{eq:SumBelief_2} to individual data streams.  
In Fig.~\ref{fig:SumStatPersonEventCamera}, we have plotted the evolution of the sum statistic for data from the camera C1. The count data is shown in Fig.~\ref{fig:CountPerCameraPersons}. The parameters learned from the data on Sept. 10th are 
$\lambda_0=0.001$, $\lambda_1=2$, $\lambda_2=4$, $\lambda_3=6$, $\lambda_4=8$, and $\lambda_5=55$. 
We have also used \eqref{eq:SumBelief_2} with $\alpha=0$.
In Fig.~\ref{fig:SumStatPersonNonEventCameras}, we have plotted the sum statistic corresponding to data from camera C1 on the non-event days, 
and from camera C2. As can be seen in the figures, the statistic corresponding to C1 on the event day fires around the event, while we see almost no activity in other streams.

\begin{figure}[ht]
	\center
	\includegraphics[width=7cm, height=5cm]{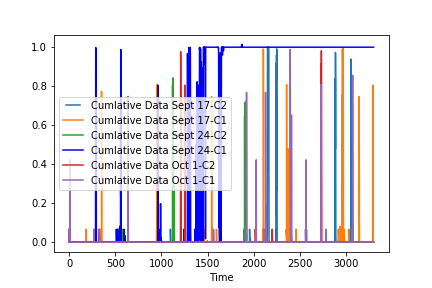}
		\caption{Evolution of the sum statistic \eqref{eq:SumBelief_2} with $\alpha=0$ applied to data from Instagram counts near camera C1 and C2 on event and non-event days.}
	\label{fig:SumStatInstagramNonEventDay}
\end{figure}

In Fig.~\ref{fig:SumStatInstagramNonEventDay}, we report results for the Instagram count data. We have again used \eqref{eq:SumBelief_2} with $\alpha=0$. The parameters learned from the data on Sept. 10th are 
$\lambda_0=0.001$, $\lambda_1=0.1$, and $\lambda_2=2$. 
The sum statistic fires and stays close to one for counts from area close to camera C1 on the event day, while there is sporadic activity for data from other streams. 
Thus, qualitatively, the sum statistic or sum belief algorithm successfully detects the 5K event. 

%
%

\section{Conclusions}
We proposed a theoretical framework for event detection in nonstationary environments using multimodal data. 
Motivated by the statistical behavior of count data extracted from CCTV images and social network posts, we formulated 
the event detection problem as a quickest change detection problem for detecting changes in count rates from one family 
of rates to another. We then obtained structural results for the optimal policy for the resulting POMDP and motivated 
a belief sum algorithm. We then applied the algorithm to real data collected around a 5K run in NYC to detect the event. 
For simplicity, we developed the framework for a single stream of count data here. However, the mathematical model can easily be extended to a vector stream of observations to detect an event jointly across modalities.
The POMDP model studied in this paper is a Bayesian model. In the future, we will explore detection in non-Bayesian models. 
We will also explore more general parametric and non-parametric models for count data for wider applicability.

\section*{Acknowledgment}
The work of Taposh Banerjee and Vahid Tarokh was supported by a grant from
the Army Research Office, W911NF-15-1-0479.

\bibliographystyle{ieeetr}
\bibliography{QCD_verSubmitted}

\end{document}